\documentclass{elsart3p}

\usepackage{bm}
\usepackage{epsfig}

\usepackage{amssymb}

\begin{document}

\begin{frontmatter}



\title{Control of test particle transport in a turbulent electrostatic model of the Scrape Off Layer}


\author{G. Ciraolo\corauthref{cor1}}
\ead{guido.ciraolo@l3m.univ-mrs.fr}
\address{MSNM-GP\thanksref{MSNMGP}, IMT La Jet\'ee, Technople de Ch\^ateau Gombert, F-13451 Marseille Cedex 20, France}
\thanks[MSNMGP]{Unit\'e Mixte de Recherche (UMR 6181) du CNRS, de l'Ecole Centrale de Marseille et des Universit\'es de Marseille.}
\author{Ph. Ghendrih,}
\author{ Y. Sarazin,}
\address{Association Euratom-CEA, DRFC/DSM/CEA, CEA Cadarache, F-13108 St. Paul-lez-Durance Cedex, France}
\author{C. Chandre,}
\author{R. Lima,}
\author{M. Vittot,}
\address{{Centre de Physique Th\'eorique\thanksref{cpt}, CNRS Luminy, Case 907, F-13288 Marseille Cedex 9, France}}
\thanks[cpt]{UMR 6207 of the CNRS, Aix-Marseille and Sud Toulon-Var Universities. Affiliated with the
CNRS Research Federation FRUMAM (FR 2291). CEA registered research laboratory LRC DSM-06-35.}
\author{M. Pettini}
\address{Instituto Nazionale di Astrofisica, Osservatorio di Arcetri, Largo Enrico Fermi 5, I-50125 Firenze, Italy,
and INFN Sezione di Firenze, Firenze, Italy}

\corauth[cor1]{Corresponding author}

\begin{abstract}
The ${\bm E}\times{\bm B}$ drift motion of charged test particle dynamics in the Scrape
Off Layer (SOL)is analyzed to investigate a transport control strategy based on
Hamiltonian dynamics. We model SOL turbulence using a 2D non-linear fluid code based on
interchange instability which was found to exhibit intermittent dynamics of the particle
flux. The effect of a small and appropriate modification of the turbulent electric
potential is studied with respect to the chaotic diffusion of test particle dynamics.
Over a significant range in the magnitude of the turbulent electrostatic field, a
three-fold reduction of the test particle diffusion coefficient is achieved.
\end{abstract}

\begin{keyword}
turbulence  \sep control \sep passive tracers \sep nonlinear dynamics

\PACS{05.45.Gg, 52.25.Fi, 52.25.Xz, 52.35.Ra, 47.27.Rc}
\end{keyword}
\end{frontmatter}

\section{Introduction}
\label{intro} Cross-field turbulent transport in the Scrape-Off-Layer (SOL) has a strong
influence on divertor efficiency and thus on the overall performance of discharges.
Recent experiments as well as numerical investigations suggest a non diffusive transport
in the SOL. The transport seems to be characterized by large intermittent events which
propagate ballistically with large velocities. On the one hand SOL turbulence can cross
H-mode barrier and connect SOL and core turbulence, destabilizing the pedestal stable
region.  On the other hand, when large intermittent events propagate in the far SOL, they
can modify the recycling pattern and consequently the wall particle content and thus the
wall tritium inventory. Furthermore, such a convection of hot ion over-densities can lead
to localised heat deposition on components that are not designed to sustain meaningful
plasma energy deposition. A strategy to control intermittent transport can therefore
prove to be highly valuable to achieve high performance operation.

Several experiments \cite{schr01,stoc05} have been dedicated to the control of turbulence
in magnetized plasmas. Although a modelling of some of the experiments has been achieved,
there was to date no theoretical backing for such a procedure. Our aim is to develop such
a theoretical background and proceed by steps towards a realistic turbulence control
scheme. In the present paper, we apply the control strategy derived for Hamiltonian
systems to the transport of test particles in an electrostatic potential which exhibits
key features of the turbulent electrostatic potential generated by the Tokam code. This
is a 2D non-linear fluid code based on interchange instability \cite{Sara98,Ghen93}.

The investigation of passive tracer dynamics has demonstrated to be a powerful tool for a
deeper understanding of fluid and plasma turbulence, even if, in the case of plasmas, one
has to be cautious about the limits of its applicability due to the back-reaction of
turbulence on electro-magnetic fields. \cite{Naul06,Naul05,Prie05,Basu03}.

Our strategy is to define a small apt modification of the system which enhances
confinement, at low additional cost of energy \cite{cira04,chan05,chan06}. In the present
case it consists in reducing the chaotic diffusion of charged test particles advected by
the ${\bm E}\times{\bm B}$ drift motion. This effect is obtained by adding a small
control term to the turbulent electric potential and comparing the chaotic diffusion of
the test particles advected by the original electric potential and by the modified one.

In the case of turbulence with intermittent bursts of transport, one has to specify the
control strategy: Either one aims at controlling turbulence in between the bursts of
transport with a fluctuation level of the order of $2$ \%, or one aims at controlling the
intermittent events with fluctuation levels larger than $10$ \%. The former is comparable
to the case of homogeneous turbulence and is readily achieved while the latter is more
challenging and is presently investigated.

The paper is organized as follows: In Section~\ref{mot_and eq} the 2D model for SOL
turbulence is recalled as well as its nonlinear transport properties. In
Section~\ref{Controllo} the procedure for the computation of the control term is
described and the results of its application to the $E\times B$ drift motion of charged
test particles are presented.

\section{Interchange turbulence in the Scrape Off Layer}
\label{mot_and eq} SOL turbulence is specific with respect to core turbulence due to the
sheath boundary conditions that strongly reduce the parallel current flowing on each flux
tube. Regimes with reduced parallel current are then very sensitive to charge separation
via the curvature drift since the parallel current is less effective to balance the
charge separation. In fact, the interchange instability, hence curvature charge
separation combined to a density gradient, in conjunction with the sheath resistivity
have been proposed to explain the very large fluctuation levels reported in the
SOL~\cite{Nedospasov89,Garbet92}. We consider here this model of SOL turbulence in the
limit $ T_i \ll T_e$. In this model two balance equations are used, the particle balance
that governs the density transport, and the charge balance which takes the form of an
evolution equation for the vorticity $\Delta \phi$, where $\phi$ is the electric
potential, due to the ion inertia terms. Further simplifications are introduced in the
flute limit, hence with weak parallel gradients such that the average along the field
lines can be performed. The system is thus reduced to a two dimensional and two field
model, $n$ the density and $\phi$ the plasma potential normalised to $T_e /e$.

\begin{eqnarray}
\left( \frac{\partial}{\partial t}-D \nabla_\perp^2 + \lbrace\phi \rbrace\right) n
  =
\sigma_{n} n  + S\nonumber\\
n \left(\frac{\partial}{\partial t} - \nu \nabla_\perp^2 + \lbrace\phi\rbrace\right)
\nabla_\perp^2 \phi + g\partial_y n  = \sigma_\phi n
  \label{syst2D}
\end{eqnarray}

In these equations time is normalised to the ion Larmor frequency $\Omega_i$ and space is
normalised to the so-called hybrid Larmor radius $\rho_s$. $D$ and $\nu$ stand for the
collisional transverse diffusion and viscosity (normalised to the Bohm values), $S$ is
the density source term, here localised radially and constant in time and along the
poloidal angle $\theta$. $\sigma_n$ and $\sigma_{\phi}$ are the sheath controlled
particle flux and current to the wall, $\sigma_n = \sigma ~exp\left(\Lambda-\phi\right)$
and $\sigma_{\phi} $=$ \sigma \big(1-exp\left(\Lambda-\phi\right)\big) $ where $\sigma=
\rho_s / qR$ stands for the normalised saturation current. The operator $\{ \phi
\}F=\{\phi,F\}=\partial_x \phi \partial_y F-\partial_y \phi
\partial_x F$ is the electric drift convection term. At first
order in the fluctuation magnitude, sheath loss terms can be linearised to yield the
Hasegawa-Wakatani coupling term $ \sigma \left(n-\phi \right)$ \cite{Hasegawa_Wakatani}.
The interested reader can refer to Refs.~\cite{Sara98,Ghen93} for more details on
Eqs.~(\ref{syst2D}), the assumptions of this model and its transport properties.

Large systems with $512 \rho_s$ in both the radial and poloidal directions are simulated
for about $200$ SOL confinement times (typically $1500$ turbulence correlation times). A
localised particle source, with Gaussian shape in the radial direction and width $8.5
\rho_s$ and no dependence in poloidal angle or time, splits the box in two regions, an
unstable region where the average density gradient is negative and exceeds the threshold
of the interchange turbulence, and a stable region with positive gradient of the average
density and where the system is stable with respect to the interchange mechanism.
\begin{figure}
\begin{center}
\epsfig{file=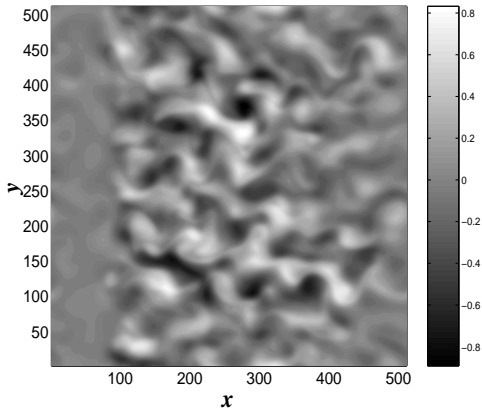,width=6.5cm,height=5.5cm}
\end{center}
\caption{Snapshot of the electric potential $\phi(x,y,t)$ for $t=100$ obtained by
Eq.~(\ref{syst2D}) on a spatial grid of $512\rho_s\times 512\rho_s$.} \label{Vphil}
\end{figure}
A contour plot of the electric potential at a given time is shown in Fig.~\ref{Vphil}.
\begin{figure}
\includegraphics[width=3.5cm,height=3.5cm]{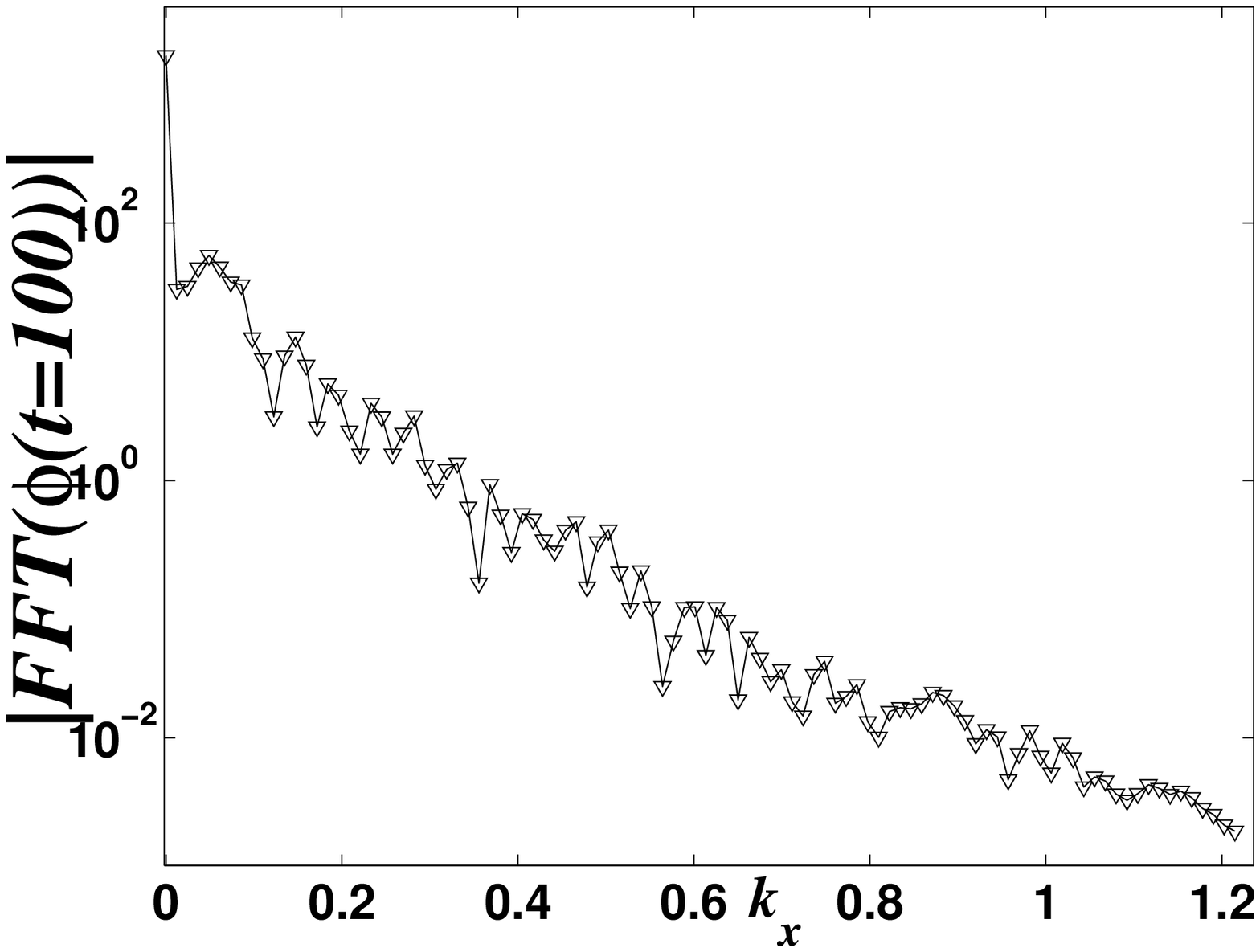}
\includegraphics[width=3.5cm,height=3.5cm]{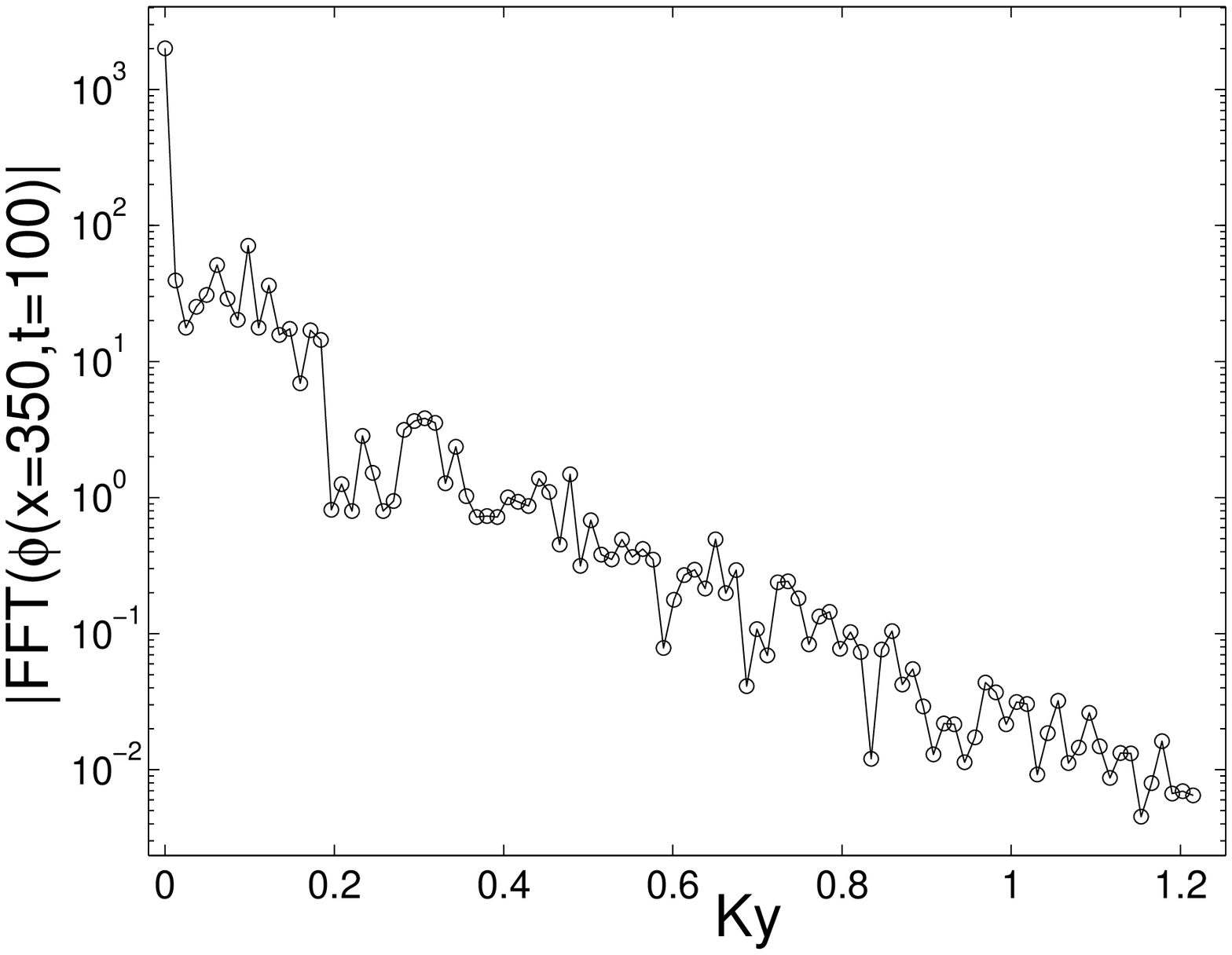}
\caption{Typical power spectra of the electric potential $\phi$ represented in
Fig.~\ref{Vphil} at $t=100$ using the same parameter as in Fig.~\ref{Vphil}.}
\label{SpettriPhi}
\end{figure}
Typical power spectra of the potential in the $x$ and $y$ direction are shown in
Fig.~\ref{SpettriPhi}.

In the following Section we study the effect of the control term on $E\times B$ drift
motion of charged test particles.

\section{Turbulent potential and test particle dynamics}
\label{Controllo} In the guiding center approximation, the equations of motion of a
charged test particle in presence of a strong toroidal magnetic field and of a time
dependent electric field are \cite{Northrop}
\begin{eqnarray}
{\dot{\bf x}}= \frac{d}{dt}{x \choose y}&=&\frac{c}{B^2}{\bf E}({\bf x},t)\times {\bf B}\nonumber\\
&=&\frac{c}{B}{-\partial_y \phi(x,y,t)\choose\partial_x \phi(x,y,t)}, \label{guidcent}
\end{eqnarray}
where $\phi$ is the electric potential, ${\bf E}=-{\bm\nabla} \phi$, and ${\bf B}=B {\bf
e}_z$. We study the 2D dynamics of test particles obeying to Eq.~(\ref{guidcent}), where
the considered turbulent potential is generated by Eqs.~(\ref{syst2D}). We consider a
reduced selected spatial region and we impose periodic boundary conditions in $x$, $y$
directions. More precisely we choose a square box of $256 \rho_s\times 256\rho_s$, not
too close to the source region (in Fig.~\ref{Vphil} the selected square box is between
$x\in[100,356]$ in the radial direction and $y\in[1,256]$ in the poloidal direction).
This allows one to avoid the complex analysis of diffusive transport in finite systems,
namely where the sources and sinks play an important role. With such an infinite system,
one can use a straightforward calculation of diffusion coefficients to characterize the
turbulent transport. Since we are interested in long time behaviour of test particle
dynamics, we impose periodic boundary conditions also in time. Periodicity in time
provides a means to generate appropriate statistics with long time series. The underlying
assumption is that all meaningful time scales of the system are shorter than the chosen
period.
\begin{figure}
\begin{center}
\epsfig{file=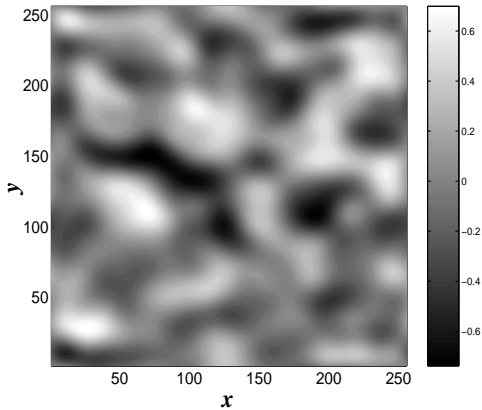,width=6.5cm,height=5.5cm}
\end{center}
\caption{Snapshot of the electric potential $V(x,y,t)$ for $t=100$ over the spatial grid
of $256\rho_s\times 256\rho_s$ obtained from the original potential $\phi$ represented in
Fig.~\ref{Vphil} after the filtering in space and time.} \label{Vrec}
\end{figure}
A snapshot of the turbulent potential in the selected spatial region after the filtering
procedure for imposing periodic boundary conditions is shown in Fig.\ref{Vrec}. In what
follows we denote $V(x,y,t)$ the filtered potential. Therefore the potential $V(x,y,t)$
is $L_x=256\rho_s$ periodic in the $x$-direction, $L_y=256\rho_s$ periodic in the
$y$-direction and $T=256\dot 50~\Omega_i^{-1}$ periodic in time.

In order to have a $1$-periodic dependence in $(x,y,t)$ for simplicity, we perform the
change of variables
\begin{equation}
\tilde x = \frac{x}{L_x},~~~\tilde y = \frac{y}{L_y},~~~\tilde t = \frac{t}{T}.
\label{resc}
\end{equation}
It follows that
\begin{equation}
\frac{d\tilde x}{d\tilde t} = \frac{T}{L_x}\frac{dx}{dt}=-\frac{T}{L_x}\frac{\partial
V}{\partial y}.
\end{equation}
From
\begin{equation}
\frac{\partial V}{\partial y}=\frac{\partial V}{\partial \tilde y}\frac{d\tilde y}{dy},
\end{equation}
we get
\begin{eqnarray}
\frac{d\tilde x}{d\tilde t}&=-&\frac{T}{L_xL_y}\frac{\partial \tilde V(\tilde x,\tilde
y,\tilde t)}{\partial \tilde y}.
\end{eqnarray}
Thus the equations in the $(\tilde x,\tilde y,\tilde t)$ variables are equivalent to the
equations in the $(x,y,t)$ variables if the potential is multiplied by a factor
$\frac{T}{L_xL_y}$.

The test particle dynamics governed by Eq.~(\ref{guidcent}) is studied with the aid of
numerical simulations. Since the potential is given over a spatio-temporal grid, the
estimation of the electric field which acts on the particle trajectories is computed
using a four point bivariate interpolation in space and a linear interpolation in time
\cite{abra70}. The time integration is performed using a fourth order Runge-Kutta scheme.

We consider a set of $\mathcal M$ particles ($\mathcal M$ of order $1000$) uniformly
distributed at random at $t=0$ and we study the diffusion properties of the system. We
verified that for the values of the parameters in the range of interest the mean square
displacement $\langle r^2 (t)\rangle$ is a linear function of time and we computed the
corresponding diffusion coefficient
$$
D^\star = \lim_{t \rightarrow\infty} \frac{\langle r^2(t)\rangle}{t},
$$
as a function of the amplitude of the turbulent potential.
\begin{figure}
\begin{center}
\epsfig{file=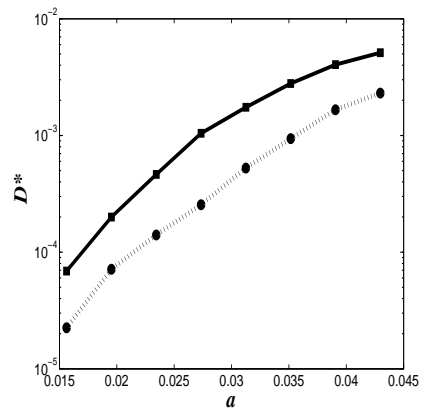,width=5.5cm,height=5.5cm}
\end{center}
\caption{Test particle diffusion coefficient $D^\star$ versus $a$ (magnitude of the
potential) in log-lin scale obtained for the turbulent potential $V$ without (full
squares, solid line) and with (full circles, dashed line) the addition of the control
term $f_2$ given by Eq.~(\ref{f2}).} \label{coeffD}
\end{figure}
In Fig.~\ref{coeffD} the results obtained for small values of the electric potential are
shown.

\begin{figure}
\begin{center}
\epsfig{file=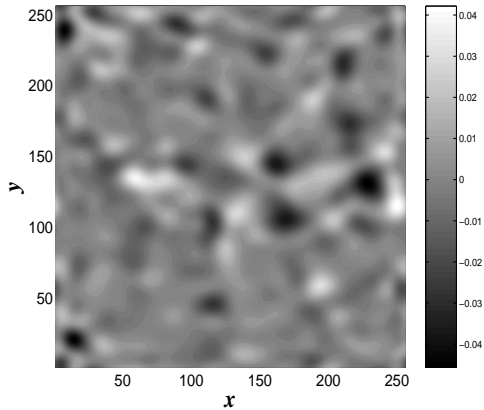,width=6.5cm,height=5.5cm}
\end{center}
\caption{Contour plot of the control term $f_2(x,y,t)$ for $t=100$ given by
Eq.~(\ref{f2}).} \label{figf2}
\end{figure}

The first step in the computation of the appropriate modification of the potential which
entails a reduction of chaotic diffusion is the Fourier expansion of the electric
potential, i.e.
\begin{equation}
V(x,y,t)=\sum_{k_1,k_2,k_3} V_{k_1k_2k_3}e^{2\pi i(k_1x+k_2y+k_3t)}. \label{Vnumerico}
\end{equation}
Then we compute $\Gamma V$, that is the time-integral of $V$,
\begin{equation}
\Gamma V=\sum_{k_1,k_2,k_3\neq0}\frac{V_{k_1k_2k_3}}{2\pi ik_3}e^{2\pi
i(k_1x+k_2y+k_3t)},
\end{equation}
and the constant part of $V$ with respect to time that is written ${\mathcal R}V$:
\begin{equation}
{\mathcal R}V=\sum_{k_1,k_2} V_{k_1k_20}~~e^{2\pi i(k_1x+k_2y)}.
\end{equation}
The control term $f_2$ is given by the Poisson bracket between $\Gamma V$ and $({\mathcal
R}+1)V$,
\begin{eqnarray}
f_2&=&-\frac{1}{2}\{\Gamma V,({\mathcal R}+1)V\}\nonumber\\
&=& -\frac{1}{2}\left(\frac{\partial\Gamma V}{\partial x}\frac{\partial({\mathcal
R}+1)V}{\partial y}-\frac{\partial\Gamma V}{\partial y}\frac{\partial({\mathcal
R}+1)V}{\partial x}\right).\nonumber\\ \label{f2}
\end{eqnarray}
A contour plot of $f_2$ at a fixed time is depicted in Fig.~\ref{figf2}.
Finally we study the test particle dynamics in the modified potential $V+f_2$. The
comparison between the diffusion coefficient obtained with and without the control term
is shown in Fig.~\ref{coeffD}. One readily observes two main features, on the one hand
the control term exhibits higher frequencies than the turbulent electrostatic field,
typically a factor $2$ larger has expected from the quadratic form of the control term,
Eq.(\ref{f2}), on the other hand the magnitude of the control term is more than a factor
$10$ smaller. This reduction compensates for the increase of the electric field via the
increased frequencies. A significant reduction of $D^\star$ is observed in the controlled
case. These preliminary results obtained for small amplitudes of the electric potential
show that the control strategy we propose can be applied to much more turbulent flows
than the one considered in \cite{cira04}. Further investigations on the efficiency domain
of the control term and its robustness are required.

\section{Discussion and conclusion}
\label{Conclusion}

In this paper, we have shown that the procedure developed to control turbulent diffusive
transport of test particles can be applied to particles transported in a realistic
turbulent electrostatic potential. The procedure to compute a low magnitude control term
provides an efficient turbulence reduction is presented. For this test of the control
strategy the turbulent field is simplified to simplify the analysis of test particle
transport. More generally, the proof of principle that a control strategy can be
efficient is tested on the simplest available model. When comparing the diffusion
coefficient of the test particles in the uncontrolled system to that in the controlled
case, a typical factor $3$ reduction of the diffusion coefficient is achieved with a
factor $10$ lower magnitude of the control term. Far more investigation is still required
on the way towards realistic control systems for the turbulent transport, however, the
present paper shows that a theory based starting point is available.

%

\end{document}